\documentclass[conference]{IEEEtran}
\IEEEoverridecommandlockouts
\usepackage{cite}
\usepackage{amsmath,amssymb,amsfonts}
\usepackage{algorithmic}
\usepackage{algorithm}
\usepackage{graphicx}
\usepackage{textcomp}
\usepackage{bm}
\usepackage{xcolor}
\usepackage{ntheorem}

\theorembodyfont{\upshape}
\theoremheaderfont{\it}
\qedsymbol{\square}
\theoremseparator{:}

\makeatletter

\newcommand{\Rmnum}[1]{\expandafter\@slowromancap\romannumeral #1@}
\makeatother
\def\BibTeX{{\rm B\kern-.05em{\sc i\kern-.025em b}\kern-.08em
    T\kern-.1667em\lower.7ex\hbox{E}\kern-.125emX}}
\begin{document}

\title{Secure Outage Analysis for RIS-Aided MISO Systems with Randomly Located Eavesdroppers
}

\author{\IEEEauthorblockN{Wei~Shi\IEEEauthorrefmark{1}\IEEEauthorrefmark{2},
Jindan~Xu\IEEEauthorrefmark{3}, 
Wei~Xu\IEEEauthorrefmark{1}\IEEEauthorrefmark{2},
Chau~Yuen\IEEEauthorrefmark{3}, A.~Lee~Swindlehurst\IEEEauthorrefmark{4}, Xiaohu~You\IEEEauthorrefmark{1}\IEEEauthorrefmark{2}, and
Chunming~Zhao\IEEEauthorrefmark{1}\IEEEauthorrefmark{2}}
\IEEEauthorblockA{\IEEEauthorrefmark{1}National Mobile Communications Research Laboratory, Southeast University, Nanjing, China}
\IEEEauthorblockA{\IEEEauthorrefmark{2}Purple Mountain Laboratories, Nanjing, China}
\IEEEauthorblockA{\IEEEauthorrefmark{3}School of Electrical and Electronics Engineering, Nanyang Technological University, Singapore}
\IEEEauthorblockA{\IEEEauthorrefmark{4}Center for Pervasive Communications and Computing, University of California, Irvine, USA}
\IEEEauthorblockA{Emails: wshi@seu.edu.cn, jindan1025@gmail.com, wxu@seu.edu.cn, chau.yuen@ntu.edu.sg, \\ swindle@uci.edu, xhyu@seu.edu.cn, cmzhao@seu.edu.cn} 
\thanks{An extended version of this article has been published in IEEE Transactions on Wireless Communications (TWC) \cite{wshi}.}
}

\maketitle

\begin{abstract}
In this paper, we consider the physical layer security of an RIS-assisted multiple-antenna communication system with randomly located eavesdroppers. The exact distributions of the received signal-to-noise-ratios (SNRs) at the legitimate user and the eavesdroppers located according to a Poisson point process (PPP) are derived, and a closed-form expression for the secrecy outage probability (SOP) is obtained. It is revealed that the secrecy performance is mainly affected by the number of RIS reflecting elements, and the impact of the transmit antennas and transmit power at the base station is marginal. In addition, when the locations of the randomly located eavesdroppers are unknown, deploying the RIS closer to the legitimate user rather than to the base station is shown to be more efficient. We also perform an analytical study demonstrating that the secrecy diversity order depends on the path loss exponent of the RIS-to-ground links. Finally, numerical simulations are conducted to verify the accuracy of these theoretical observations.
\end{abstract}

\section{Introduction}
Reconfigurable intelligent surface (RIS) technology has~recently been recognized as a promising approach for realizing both spectral and energy efficient communications in future wireless networks \cite{1,2,200}. An RIS comprises a large number of low-cost passive reflecting elements that are able to independently control the phases and/or amplitudes of their reflection coefficients. Due to their reconfigurable behavior, RISs have~been widely considered for various wireless applications \cite{103,104,106,107-1,107-2,119}.

In recent years, security for wireless communication has become a critical issue \cite{123}\cite{54250}. The capability of RIS to create a smart controllable wireless propagation makes it a promising approach for providing physical layer security (PLS) \cite{120}. There are multiple works that investigate the theoretical secrecy performance for RIS-enhanced PLS systems \cite{3,4,5}. However, for analytical simplicity and mathematical tractability, most work has considered single-antenna nodes and Rayleigh fading channels, and overlooked randomly distributed eavesdropper locations. Although the authors of \cite{4} and \cite{5} considered the random eavesdropper locations, there are still several research gaps left to be filled. In \cite{4}, Rician fading channels and optimization of the RIS phase shifts were not taken into consideration for the considered multiple-antenna scenario. In \cite{5}, the study was conducted based on a simplified transmit beamforming design and a secrecy diversity order analysis was not conducted.  

In this paper, we investigate the secrecy performance of an RIS-assisted multiple-input single-output (MISO) system with randomly located eavesdroppers. We first derive the exact distributions of the received signal-to-noise-ratios (SNRs) for the legitimate user and the eavesdroppers. Then, we present a closed-form expression for the secrecy outage probability (SOP). The obtained expression shows that the SOP is mainly affected by the number of RIS reflecting elements, and is not a strong function of the number of transmit antennas nor the transmit power at the base station. In addition, when the locations of the randomly located eavesdroppers are unknown, it is shown that deploying the RIS closer to the legitimate user is more efficient. To obtain more insightful observations, an asymptotic SOP analysis at high SNR is also conducted. It is shown that the secrecy diversity order ultimately only depends on the path loss exponent of the RIS-to-ground links.


\section{System Model}
\begin{figure}[!t]
\centering
\includegraphics[width=2.7in]{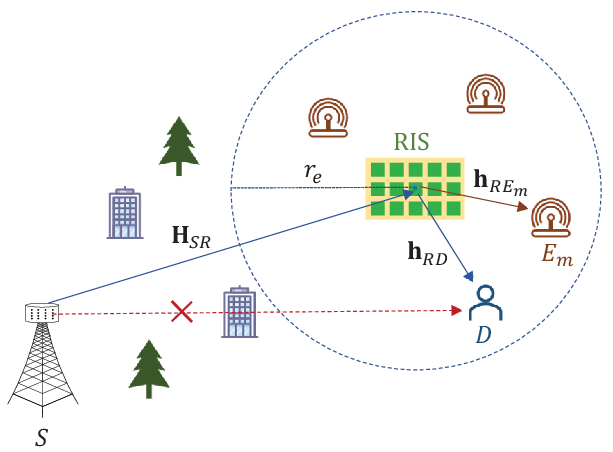}
\caption{System model in the presence of randomly located eavesdroppers.}
\label{fig1}
\end{figure}

We consider an RIS-assisted secure communication system consisting of a base station ($S$) with $K$ antennas and an RIS with $N$ reflecting elements, as illustrated in Fig.~\ref{fig1}. The reflection coefficient matrix of the RIS is denoted by $\mathbf{\Theta}\triangleq {\rm diag}\left\{{\eta_1}{\rm e}^{j\theta_1},\ldots,{\eta_n}{\rm e}^{j\theta_n},\ldots,{\eta_{N}}{\rm e}^{j\theta_N}\right\}$, where ${\rm diag}\left\{\cdot\right\}$ indicates a diagonal matrix, and $\theta_n\in[0,2\pi)$ ($\eta_n\in[0,1]$) is the phase (amplitude) coefficient of the $n$-th reflecting element. In order to exploit the maximum reflection capability of the RIS, the amplitude coefficients in this work are set to 1, i.e., $\eta_{n}\!\!=\!\!1$ for all $n$. The spatial distribution of the randomly located eavesdroppers ($E$) in a disk of radius $r_e$ centered at the RIS is modeled using a homogeneous Poisson point process (PPP), which is denoted by $\Phi_e$ with a density $\lambda_{e}$, while the legitimate user can locate randomly without the restriction of this disk.

We assume that the direct link between $S$ and the legitimate user ($D$) is blocked by obstacles. In this scenario, the data transmission between $S$ and $D$ is ensured by the RIS. Since the base station and RIS are usually deployed at an elevated height, the channel between $S$ and the RIS can be assumed to be line-of-sight (LoS) \cite{116-1}\cite{116-2}, denoted by $\mathbf{H}_{S\!R}\in\mathbb{C}^{N\times K}$. While the legitimate user and eavesdroppers are usually located on the ground, the RIS-related channels with these terminals undergo both direct LoS and rich scattering, which can be modeled using Rician fading. Here, $\mathbf{h}_{Ri}\in\mathbb{C}^{N\times1}$ is the channel vector of the RIS-$i$ links, where $i\in\left\{D,E_m\right\}$ and $E_m$ represents the $m$-th eavesdropper. Specifically, the expressions for $\mathbf{H}_{S\!R}$ and $\mathbf{h}_{Ri}$ are given by
\begin{align}
\mathbf{H}_{S\!R}\!=\!\sqrt\nu {\overline{\mathbf{H}}}_{S\!R}, \mathbf{h}_{Ri}\!=\!\!\sqrt{\mu_i}\!\left(\sqrt{\frac{\epsilon}{\epsilon\!+\!1}}{\overline{\mathbf{h}}}_{Ri}\!+\!\sqrt{\frac{1}{\epsilon\!+\!1}}{\widetilde{\mathbf{h}}}_{Ri}\!\right),
\label{equal1}
\end{align}
where $\nu=\beta_0d_{S\!R}^{-\alpha_1}$ and $\mu_i=\beta_0d_{Ri}^{-\alpha_2}$ denote the large-scale fading coefficients, $\beta_0$ is the path loss at a reference distance of $1$m, $d_{S\!R}$ ($d_{Ri}$) and $\alpha_1$ ($\alpha_2$) are the distances and the path loss exponents of the $S$-RIS (RIS-$i$) links respectively, and $\epsilon$ denotes the Rician factor. The vector ${\widetilde{\mathbf{h}}}_{Ri}$ represents the non-line-of-sight (NLoS) component, whose entries are standard independent and identically distributed (i.i.d.) Gaussian random variables (RVs). The LoS components ${\overline{\mathbf{H}}}_{S\!R}$ and ${\overline{\mathbf{h}}}_{Ri}$ are expressed as
\begin{align}
{\overline{\mathbf{H}}}_{S\!R}=\mathbf{a}_N\left(\phi_{SR}^a,\phi_{SR}^e\right)\mathbf{a}_K^H\left(\psi_{SR}^a,\psi_{SR}^e\right)=\mathbf{a}_{N,SR}\mathbf{a}_{K,SR}^H,
\label{equal3}
\end{align}
\begin{align}
{\overline{\mathbf{h}}}_{Ri}=\mathbf{a}_N\left(\psi_{Ri}^a,\psi_{Ri}^e\right)=\mathbf{a}_{N,Ri},
\label{equal4}
\end{align}
where $\phi_{S\!R}^a$ ($\phi_{S\!R}^e$) is the azimuth (elevation) angle of arrival (AoA) at the RIS, $\psi_{S\!R}^a$ ($\psi_{S\!R}^e$) and $\psi_{Ri}^a$ ($\psi_{Ri}^e$) are the azimuth (elevation) angles of departure (AoD) at the base station and RIS, respectively, and $\mathbf{a}_Z\left(\vartheta^a,\vartheta^e\right)$ is the array response vector expressed as \cite{6}
\begin{align}
\mathbf{a}_Z\left(\vartheta^a,\vartheta^e\right)&\!=\!\left[1,\ldots,{\rm e}^{j2\pi\frac{d}{\lambda}\left(x\sin{\vartheta^a}\sin{\vartheta^e}+y\cos{\vartheta^e}\right)},\ldots,\right.\nonumber\\
&\left.{\rm e}^{j2\pi\frac{d}{\lambda}\left(\left(\sqrt Z\!-\!1\right)\sin{\vartheta^a}\sin{\vartheta^e}+\left(\sqrt Z\!-\!1\right)\cos{\vartheta^e}\right)}\right]^T,
\label{equal5}
\end{align}
where $d$ and $\lambda$ are the element spacing and signal wavelength, and $0\le x,y<\sqrt Z$ are the element indices in the plane.

\section{Distributions of the Received SNRs}
In order to analyze the secrecy performance of the system, we need to first characterize the distributions of the received SNRs at the legitimate user and the eavesdroppers.
\subsection{Distribution of the Received SNR at $D$}
Assuming quasi-static flat fading channels, the signal received at $D$ is expressed as
\begin{equation}
r_D=\mathbf{g}_D^H\mathbf{f}s+n_D,
\label{equal6}
\end{equation}
where $\mathbf{g}_D^H\triangleq\mathbf{h}_{R\!D}^H\mathbf{\Theta}\mathbf{H}_{S\!R}$ denotes the cascaded channel, $\mathbf{f}$ is the normalized beamforming vector, $s$ denotes the transmit signal that satisfies the power constraint $\mathbb{E}\{\left|s\right|^2\}=P_T$, $\mathbb{E}\{\cdot\}$ is the expectation of a RV, and $n_D\sim\mathcal{CN}(0,\sigma_D^2)$ is the additive white Gaussian noise (AWGN) at $D$ with variance $\sigma_D^2$, where ${\cal {CN}}$ is the complex Gaussian distribution. 
Therefore, the received SNR at $D$ is calculated as
\begin{equation}
\gamma_D=\frac{P_T\left|\mathbf{g}_D^H\mathbf{f}\right|^2}{\sigma_D^2}=\rho_d\left|{\rm A}\right|^2,
\label{equal7}
\end{equation}
where $\left|{\rm A}\right|\triangleq\left|\mathbf{g}_D^H\mathbf{f}\right|$, and $\rho_d\triangleq\frac{P_T}{\sigma_D^2}$ denotes the transmit SNR.

\emph{Theorem~1:} When MRT beamforming is adopted, i.e., $\mathbf{f}=\frac{\mathbf{g}_D}{\left\|\mathbf{g}_D^H\right\|}$, the optimal phase shift matrix of the RIS is given as
\begin{equation}
\mathbf{\Theta}^{\star}={\rm diag}\left\{{\rm e}^{-j\angle\left({\rm diag}\left\{\mathbf{h}_{R\!D}^H\right\}\mathbf{a}_{N\!,SR}\right)}\right\},
\label{equal8}
\end{equation}
where $\angle$ returns the phase of a complex value.

\emph{Proof:} See Appendix A.$\hfill\blacksquare$

With the optimized RIS phase shifts in \emph{Theorem~1}, the RV $\left|{\rm A}\right|$ is expressed as $\left|{\rm A}\right|\!=\!\sqrt{K\nu}\sum_{n=1}^{N}\left|h_{R\!D}\left(n\right)\right|$ which follows the distribution characterization in the following lemma.

\emph{Lemma~1:} The cumulative distribution function (CDF) of $\left|{\rm A}\right|$ is well approximated by
\begin{equation}
F_{\left|{\rm A}\right|}\left(x\right)=\frac{1}{\Gamma\left(k\right)}\gamma\left(k,\frac{x}{\theta}\right),
\label{equal9}
\end{equation}
where $\Gamma\left(\cdot\right)$ is the Gamma function, $\gamma\left(\cdot,\cdot\right)$ denotes the lower incomplete Gamma function \cite[Eq.~(8.350.1)]{10}, with shape parameter $k=N\frac{\frac{\pi}{4}\left(L_\frac{1}{2}\left(-\epsilon\right)\right)^2}{1+\epsilon-\frac{\pi}{4}\left(L_\frac{1}{2}\left(-\epsilon\right)\right)^2}$ and scale parameter $\theta=\sqrt K\sqrt{\frac{\mu_D\nu}{\epsilon+1}}\frac{1+\epsilon-\frac{\pi}{4}\left(L_\frac{1}{2}\left(-\epsilon\right)\right)^2}{\frac{\sqrt\pi}{2}L_\frac{1}{2}\left(-\epsilon\right)}$, in which $L_q\left(x\right)$ is the Laguerre polynomial defined in \cite[Eq.~(2.66)]{7}.

\emph{Proof:} See Appendix B.$\hfill\blacksquare$

By applying \emph{Lemma~1}, we can obtain the CDF and probability density function (PDF) of $\gamma_{D}$, respectively, as
\begin{equation}
F_{\gamma_D}\left(x\right)=F_{\left|{\rm A}\right|}\left(\sqrt{{x}/{\rho_d}}\right)=\frac{1}{\Gamma\left(k\right)}\gamma\left(k,\frac{\sqrt{{x}/{\rho_d}}}{\theta}\right),
\label{equal10}
\end{equation}
and
\begin{equation}
f_{\gamma_D}\left(x\right)=\frac{dF_{\gamma_D}\left(x\right)}{dx}=\frac{{\rm e}^{-\frac{\sqrt{{x}/{\rho_d}}}{\theta}}\left(\frac{\sqrt{{x}/{\rho_d}}}{\theta}\right)^k}{2\Gamma\left(k\right)x}.
\label{equal11}
\end{equation}

\subsection{Distribution of the Received SNR at $E$}
Before calculating the effective SNR of the independent and homogeneous PPP distributed eavesdroppers, we first derive the SNR of the $m$-th eavesdropper $E_m$. The signal received at $E_m$ is formulated as
\begin{equation}
r_{E_m}=\mathbf{h}_{R\!E_m}^H\mathbf{\Theta}\mathbf{H}_{S\!R}\mathbf{f}s+n_{E_m},
\label{equal12}
\end{equation}
where $n_{E_m}\sim\mathcal{CN}(0,\sigma_E^2)$ is AWGN at $E_m$ with variance $\sigma_E^2$.
The received SNR at $E_m$ is given as follows.

\emph{Proposition~1:} The received SNR at $E_m$ is expressed as
\begin{equation}
\gamma_{E_m}=\rho_e \left|\mathbf{h}_{R\!E_m}^H\mathbf{\Theta}\mathbf{H}_{S\!R}\mathbf{f}\right|^2=\rho_eK\nu\left|Z_{E_m}\right|^2,
\label{equal13}
\end{equation}
where $\rho_e\triangleq\frac{P_T}{\sigma_E^2}$ denotes the transmit SNR and we define the RV $Z_{E_m}\!\triangleq\!\sum_{n=1}^{N}{h_{R\!E_m}^\ast\left(n\right){\rm e}^{-j\angle h_{R\!D}^\ast\left(n\right)}}$.

According to \emph{Proposition~1}, we present \emph{Lemma~2} before deriving the distribution of $\gamma_{E_m}$.

\emph{Lemma~2:} The RV $Z_{E_m}$ follows a complex Gaussian distribution with mean $M_{E_m}$ and variance $V_{E_m}$, where $M_{E_m}=\sqrt{\frac{\mu_{E_m}\epsilon^2}{{\frac{\pi}{4}\left(\epsilon+1\right)\left(\!L_\frac{1}{2}\!\left(-\epsilon\right)\!\right)}^2}}{\rm e}^{j\pi\frac{d}{\lambda}\!\left(\!\sqrt N\!-\!1\!\right)\left(\!\delta_1\!+\!\delta_2\!\right)}\frac{\sin{\left(\!\pi\frac{d}{\lambda}\sqrt N \delta_1\!\right)}\sin{\left(\!\pi\frac{d}{\lambda}\sqrt N \delta_2\!\right)}}{\sin{\left(\pi\frac{d}{\lambda}\delta_1\right)}\sin{\left(\pi\frac{d}{\lambda}\delta_2\right)}}$, $V_{E_m}=N\mu_{E_m}\!\left[1\!-\!\frac{\epsilon^2}{\frac{\pi}{4}\left(\epsilon+1\right){\left(L_\frac{1}{2}\left(-\epsilon\right)\right)}^2}\right]$, $\delta_1\!=\!\sin{\psi_{R\!D}^a}\sin{\psi_{R\!D}^e}\!\\-\!\sin{\psi_{R\!E_m}^a}\sin{\psi_{R\!E_m}^e}$ and $\delta_2\!=\!\cos{\psi_{R\!D}^e}\!-\!\cos{\psi_{R\!E_m}^e}$.

\emph{Proof:} See Appendix C.$\hfill\blacksquare$

As disclosed in \emph{Lemma~2}, we conclude that $\gamma_{E_m}$ is a non-central Chi-squared RV with two degrees of freedom. Then, the CDF of $\gamma_{E_m}$ is given by
\begin{equation}
F_{\gamma_{E_m}}\left(x\right)=1-Q_1\left(\frac{s}{\sigma},\frac{\sqrt x}{\sigma}\right),
\label{equal14}
\end{equation}
where $s\!=\!\sqrt{\frac{\rho_eK\nu\mu_{E_m}\epsilon^2}{\frac{\pi}{4}\left(\epsilon+1\right){\left(L_\frac{1}{2}\left(-\epsilon\right)\right)}^2}}\left|\frac{\sin{\left(\pi\frac{d}{\lambda}\sqrt N \delta_1\right)}\sin{\left(\pi\frac{d}{\lambda}\sqrt N \delta_2\right)}}{\sin{\left(\pi\frac{d}{\lambda}\delta_1\right)}\sin{\left(\pi\frac{d}{\lambda}\delta_2\right)}}\right|$, $\sigma^2\!=\!\frac{1}{2}\rho_eKN\nu\mu_{E_m}\left[1-\frac{\epsilon^2}{\frac{\pi}{4}\left(\epsilon+1\right){\left(L_\frac{1}{2}\left(-\epsilon\right)\right)}^2}\right]$, and $Q_1({\rm a},{\rm b})$ is the first-order Marcum $Q$-function \cite{9}.

In the case of non-colluding eavesdroppers, the eavesdropper with the strongest channel dominates the secrecy performance. Thus, the corresponding CDF of the eavesdropper SNR is derived as
\begin{align}
F_{\gamma_E}\left(x\right)&={\rm Pr}\left\{\mathop{\max}\limits_{m\in{\Phi_e}}{\gamma_{E_m}\le x}\right\}\nonumber\\
&\mathop=^{\left(\rm a\right)}\mathbb{E}_{\Phi_e}\left\{\prod_{m\in\Phi_e,r_m\le r_e}{F_{\gamma_{E_m}}\left(x\right)}\right\}\nonumber\\
&\mathop=^{\left(\rm b\right)}{\rm exp}\left[-2\pi\lambda_e\int_{0}^{r_e}{\left(1-F_{\gamma_{E_m}}\left(x\right)\right)r\,{\rm{d}}r}\right]\nonumber\\
&\mathop=^{\left(\rm c\right)}{\rm exp}\left[-2\pi\lambda_e\int_{0}^{r_e}{Q_1\left(\varpi,\Xi\sqrt x r^\frac{\alpha_2}{2}\right)r\,{\rm{d}}r}\right],
\label{equal15}
\end{align}
where $({\rm a})$ follows from the i.i.d. characteristic of the eavesdroppers' SNRs and their independence from the point process ${\Phi_e}$, $({\rm b})$ follows from the probability generating functional (PGFL) of the PPP \cite[Eq.~(4.55)]{8}, and $({\rm c})$ is obtained by using $\mu_{E_m}\!\!=\!\!\beta_0r^{-\alpha_2}$ and defining $\varpi\triangleq\sqrt2\!\left|\!\frac{\sin{\left(\pi\frac{d}{\lambda}\sqrt N \delta_1\right)}\sin{\left(\pi\frac{d}{\lambda}\sqrt N \delta_2\right)}}{\sin{\left(\pi\frac{d}{\lambda}\delta_1\right)}\sin{\left(\pi\frac{d}{\lambda}\delta_2\right)}}\!\right|\!\left[N\!\!\left(\frac{\frac{\pi}{4}\left(\epsilon+1\right){\left(L_{1/2}\left(-\epsilon\right)\right)}^2}{\epsilon^2}\!-\!1\right)\right]^{-\frac{1}{2}}$ and $\Xi\triangleq{\sqrt2}\left(NK\nu\beta_0\rho_e\left[1-\frac{\epsilon^2}{\frac{\pi}{4}\left(\epsilon+1\right){\left(L_{1/2}\left(-\epsilon\right)\right)}^2}\right]\right)^{-\frac{1}{2}}$.

From the characterization in \cite[Eq.~(2)]{9}, we have a tight approximation for the Marcum $Q$-function in (\ref{equal15}), that is, $Q_1\left(\varpi,\Xi\sqrt x r^\frac{\alpha_2}{2}\right)\simeq{\rm exp}\left[-{\rm e}^{v\left(\varpi\right)}\left(\Xi\sqrt x r^\frac{\alpha_2}{2}\right)^{\mu\left(\varpi\right)}\right]$, where $v\left(\varpi\right)$ and $\mu\left(\varpi\right)$ are polynomial functions of $\varpi$ defined as $v\left(\varpi\right)\!=\!-0.840\!+\!0.327\varpi\!-\!0.740\varpi^2\!+\!0.083\varpi^3\!-\!0.004\varpi^4$ and $\mu\left(\varpi\right)\!=\!2.174\!-\!0.592\varpi\!+\!0.593\varpi^2\!-\!0.092\varpi^3\!+\!0.005\varpi^4$. Then, (\ref{equal15}) is further calculated as
\begin{align}
F_{\gamma_E}\left(x\right)&\!=\!{\rm exp}\left[\!-\!2\pi\lambda_e\!\int_{0}^{r_e}\!{{\rm exp}\!\left[-{\rm e}^{v\left(\varpi\right)}\left(\Xi\sqrt x r^\frac{\alpha_2}{2}\right)^{\mu\left(\varpi\right)}\right]\!r{\rm{d}}r}\right]\nonumber\\
&\!=\!{\rm exp}\left[\!-t_0\frac{\Gamma\left(t_1\right)-\Gamma\left(t_1,t_2x^{t_3}\right)}{x^{t_4}}\right],
\label{equal16}
\end{align}
where the last equality is obtained from \cite[Eq.~(3.326)]{10} with the definitions $t_0=\frac{2\pi\lambda_e}{\frac{\alpha_2}{2}\mu\left(\varpi\right){\rm e}^\frac{4v\left(\varpi\right)}{\alpha_2\mu\left(\varpi\right)}\Xi^\frac{4}{\alpha_2}}$, $t_1=\frac{2}{\frac{\alpha_2}{2}\mu\left(\varpi\right)}$, $t_2={\rm e}^{v\left(\varpi\right)}\Xi^{\mu\left(\varpi\right)}{r_e}^{\frac{\alpha_2}{2}\mu\left(\varpi\right)}$, $t_3=\frac{\mu\left(\varpi\right)}{2}$, and $t_4=\frac{2}{\alpha_2}$.


Therefore, the PDF of the overall eavesdropper SNR could be further derived from (\ref{equal16}) as
\begin{align}
f_{\gamma_E}\left(x\right)=&\frac{dF_{\gamma_E}\left(x\right)}{dx}=t_0x^{-t_4-1}\left(t_4\gamma\left(t_1,t_2x^{t_3}\right)\right.\nonumber\\
&\left.-t_3\left(t_2x^{t_3}\right)^{t_1}{\rm e}^{-t_2x^{t_3}}\right){\rm e}^{-t_0x^{-t_4}\gamma\left(t_1,t_2x^{t_3}\right)}.
\label{equal17}
\end{align}

\section{Secrecy Outage Analysis}
In this section, we apply the derived statistical properties of $\gamma_D$ and $\gamma_E$ in the above section section to conduct the secrecy outage analysis of the RIS-aided MISO system. 
\subsection{Theoretical SOP Analysis}
A popular metric for quantifying the PLS is the SOP, which is defined as the probability that the instantaneous secrecy capacity falls below a target secrecy rate $C_{\rm th}$. Mathematically, the SOP is evaluated by
\begin{align}
{\rm SOP}&={\rm Pr}\left(\ln{\left(1+\gamma_D\right)}-\ln{\left(1+\gamma_E\right)}<C_{\rm th}\right)\nonumber\\
&=\int_{0}^{\infty}{F_{\gamma_D}\left(\left(1+x\right)\varphi-1\right)}f_{\gamma_E}\left(x\right){\rm{d}}x,
\label{equal18}
\end{align}
where $\varphi\triangleq {\rm e}^{C_{\rm th}}$. 

In order to analyze the secrecy performance, a closed-form expression for the SOP is presented in \emph{Proposition~2}.

\emph{Proposition~2:} When $r_e\rightarrow\infty$, the SOP can be approximated by the following
\begin{align}
{\rm SOP}\simeq~&1-\frac{1}{\Gamma\left(k\right)}\frac{p^\frac{1}{2}q^{k-\frac{1}{2}}}{2^{\frac{p+4q}{2}-2k}\pi^{\frac{p+4q}{2}-1}}\nonumber\\
&\times G_{0,p+4q}^{p+4q,0}\left(\frac{\left(t_0\Gamma\left(t_1\right)\varphi^{t_4}\right)^p}{p^p\left(4q\sqrt{\rho_d}\theta\right)^{4q}}\middle|\begin{matrix}-\\\Delta\\\end{matrix}\right),
\label{equal22}
\end{align}
where $G_{s,t}^{m,n}\left(z\right)$ is Meijer's $G$ function \cite{10}, $p,q\in\mathbb{Z}^+$, ${p}/{q}=\alpha_2$, and $\Delta=\left[0,\frac{1}{p},\ldots,\frac{p-1}{p},\frac{k}{4q},\frac{k+1}{4q},\ldots,\frac{k+4q-1}{4q}\right]$.

\emph{Proof:} See Appendix D.$\hfill\blacksquare$

A number of interesting points can be noted from (\ref{equal22}).

\emph{Remark~1:} From (\ref{equal22}), we see that the transmit power $P_T$ affects only the term $\frac{t_0^p}{\rho_d^{2q}}\!\!\propto\!\!\left(\frac{\rho_e}{\rho_d}\right)^{2q}$. Thus, we see that the SOP is not a function of $P_T$, and increasing the transmit power does not improve the secrecy performance. This is intuitive since an increase in $P_T$ yields a proportional increase in both the transmit SNRs at the legitimate user and the eavesdroppers.

\emph{Remark~2:} Acoording to (\ref{equal22}), we obtain that $\frac{t_0^p}{\theta^{4q}}\!\!\propto\!\!\left(\Xi\theta\right)^{-4q}$, where $\Xi\theta\!\!=\!\!\frac{\chi}{\sqrt N}$, and the coefficient $\chi$ is independent of $N$ and $K$. This implies that the SOP is mainly affected by the number of RIS reflecting elements, $N$, and the impact of the number of transmit antennas, $K$, is marginal. 

In addition, we see that \emph{Proposition~2} is the general analysis for any rational path loss exponent. Some specific case studies are reported as follows.

\emph{Corollary~1:} For the special case of $\alpha_2=2$, i.e., $p=2$ and $q=1$, which corresponds to free space propagation \cite{124}, the SOP in (\ref{equal22}) reduces to
\begin{align}
{\rm SOP}\simeq1-\frac{2^{k-1}}{\sqrt\pi\Gamma\left(k\right)}\ G_{0,3}^{3,0}\left(\frac{t_0\Gamma\left(t_1\right)\varphi}{4\rho_d\theta^2}\middle|\begin{matrix}-\\0,\frac{k}{2},\frac{k+1}{2}\\\end{matrix}\right).
\label{equal23}
\end{align}

\emph{Corollary~2:} For the special case of $\alpha_2=4$, i.e., $p=4$ and $q=1$, which is a common practical value for the path-loss exponent in outdoor urban environments \cite{124}, the SOP in (\ref{equal22}) simplifies to the following expression
\begin{align}
{\rm SOP}\simeq1\!-\!\frac{2}{\Gamma\left(k\right)}\!\left(\frac{t_0\Gamma\left(t_1\right)\sqrt\varphi}{\sqrt{\rho_d}\theta}\right)^\frac{k}{2}\!\!K_k\!\!\left(\!2\!\left(\frac{t_0\Gamma\left(t_1\right)\sqrt\varphi}{\sqrt{\rho_d}\theta}\right)^\frac{1}{2}\!\right),
\label{equal23-1}
\end{align}
where $K_\nu\left(\cdot\right)$ denotes the $\nu$-th-order modified Bessel function of the second kind \cite[Eq.~(8.407)]{10}.

\emph{Remark~3:} From (\ref{equal23-1}), we obtain that the SOP is a monotonically increasing function w.r.t. $\frac{t_0}{\sqrt{\rho_d}\theta}=\sqrt{\frac{\rho_e}{\rho_d}}\lambda_ed_{R\!D}^2\beta\left(N,\epsilon\right)$ with fixed $k$, where $\beta\left(N,\epsilon\right)$ consists of parameters $N$, $\epsilon$, and constant terms. We note that the SOP increases with the density parameter $\lambda_e$, which implies that a larger density of randomly located eavesdroppers leads to a negative effect on the secrecy performance. Moreover, we can also see that the SOP is only related to the distance of the RIS-$D$ link, i.e., $d_{R\!D}$. Therefore, when the locations of the eavesdroppers are unknown, this suggests that the RIS should be deployed closer to the legitimate user than to the base station.

\subsection{Secrecy Diversity Order Analysis}
In order to derive the secrecy diversity order and gain further insights, we adopt the analytical framework proposed in \cite{11} where the secrecy diversity order is defined as follows
\begin{align}
d_s=-\lim_{\rho_d\rightarrow\infty}{\frac{\log{{\rm SOP}^\infty}}{\log{\rho_d}}},
\label{equal24}
\end{align}
where ${\rm SOP}^\infty$ represents the asymptotic value of the SOP in (\ref{equal22}) for $\rho_d\rightarrow\infty$, and the transmit SNR $\rho_e$ is set to arbitrary fixed values.

According to \cite[Eq.~(07.34.06.0006.01)]{16}, the SOP in (\ref{equal22}) can be expanded as
\begin{align}
{\rm SOP}\simeq&1-\frac{1}{\Gamma\left(k\right)}\frac{p^\frac{1}{2}q^{k-\frac{1}{2}}}{2^{\frac{p+4q}{2}-2k}\pi^{\frac{p+4q}{2}-1}}\times G_{0,p+4q}^{p+4q,0}\left(x\middle|\begin{matrix}-\\\Delta\\\end{matrix}\right) \nonumber\\
=&1-\frac{1}{\Gamma\left(k\right)}\frac{p^\frac{1}{2}q^{k-\frac{1}{2}}}{2^{\frac{p+4q}{2}-2k}\pi^{\frac{p+4q}{2}-1}}\times\nonumber\\
&\sum_{l=1}^{p+4q}{\prod_{j=1,j\neq l}^{p+4q}\!\Gamma\left(\Delta\left(j\right)\!-\!\Delta\left(l\right)\right)x^{\Delta\left(l\right)}}\left(1\!+\!\mathcal{O}\left(x\right)\right),
\label{equal25}
\end{align}
where $x\!=\!\!\frac{\left(t_0\Gamma\left(t_1\right)\varphi^{t_4}\right)^p}{p^p\left(4q\sqrt{\rho_d}\theta\right)^{4q}}\!\rightarrow\!0$, and $\mathcal{O}$ denotes higher order terms. 

When the transmit SNR $\rho_d\rightarrow\infty$, only the dominant terms $l=0$ and $l=1$ in the summation of (\ref{equal25}) are retained, which yields the asymptotic SOP as
\begin{align}
{\rm SOP}^\infty=&1-\frac{1}{\Gamma\left(k\right)}\frac{p^\frac{1}{2}q^{k-\frac{1}{2}}}{2^{\frac{p+4q}{2}-2k}\pi^{\frac{p+4q}{2}-1}}\left[\prod_{j=2}^{p+4q}\Gamma\left(\Delta\left(j\right)\right)x^0\right.\nonumber\\
&\left.+\prod_{j=1,j\neq2}^{p+4q}\Gamma\left(\Delta\left(j\right)-\frac{1}{p}\right)x^\frac{1}{p}\right] \nonumber\\
=&\frac{t_0\mathrm{\Gamma}\left(t_1\right)\varphi^\frac{2}{\alpha_2}\Gamma\left(k-\frac{4}{\alpha_2}\right)}{\theta^\frac{4}{\alpha_2}\Gamma\left(k\right)}\left(\rho_d\right)^{-\frac{2}{\alpha_2}},
\label{equal27}
\end{align}
where the last step is calculated by applying Gauss' multiplication formula \cite[Eq.~(6.1.20)]{19}.
 
\emph{Remark~4:} By substituting (\ref{equal27}) into (\ref{equal24}), the secrecy diversity order is obtained as $\frac{2}{\alpha_2}$, which only depends on the path loss exponent of the RIS-to-ground links. This implies that the secrecy diversity order of this system improves when the RIS is deployed to provide better LoS links to the terminals.

\section{Simulation Results}

\begin{figure}[!t]
    \setlength{\abovecaptionskip}{0pt}
    \setlength{\belowcaptionskip}{0pt}
    \centering
    \includegraphics[width=2.8in]{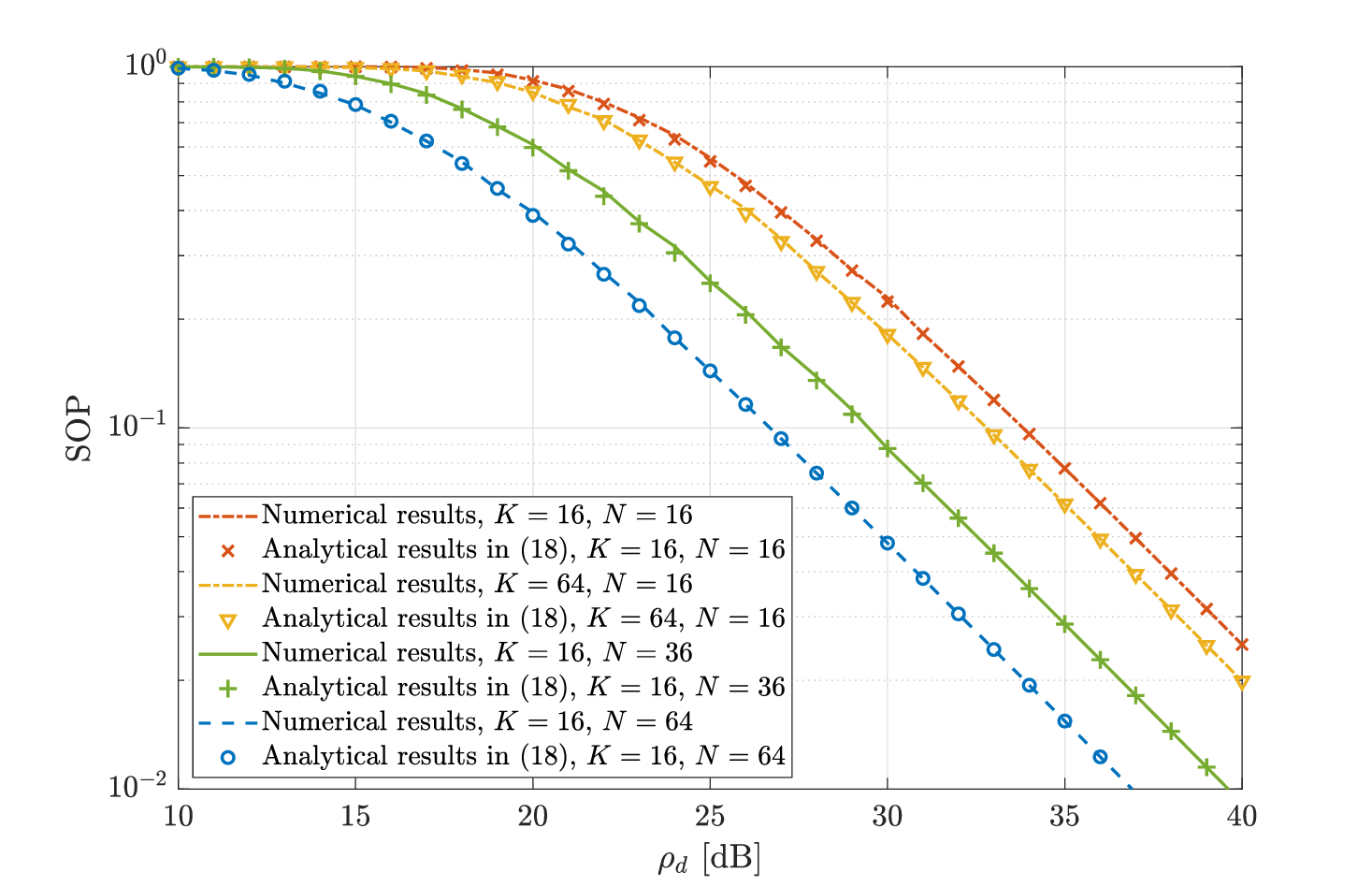}
    \caption{The SOP versus $\rho_d$, with $\alpha_1=\alpha_2=2$, $\epsilon=2$, $d_{S\!R}=30~{\rm m}$, $d_{R\!D}=40~{\rm m}$, $r_e=200~{\rm m}$, $\lambda_e=10^{-3}$, $C_{\rm th}=0.05$, and $\rho_e= 30~{\rm dB}$.}
    \vspace{-0.45cm}
    \label{fig3} \end{figure}

In this section, Monte-Carlo simulations are illustrated to validate the analytical results.
Fig.~\ref{fig3} depicts the SOP versus the transmit SNR $\rho_d$ for different values of $N$ and $K$. The analytical expressions in (\ref{equal22}) match very well with the numerical results. Furthermore, as expected from \emph{Remark~2}, the SOP obviously decreases as $N$ increases. However, the SOP remains almost the same when $K$ increases with fixed $N$, which means that the impact of the number of transmit antennas on the secrecy performance is negligible.

\begin{figure}[!t]
    \setlength{\abovecaptionskip}{0pt}
    \setlength{\belowcaptionskip}{0pt}
    \centering
    \includegraphics[width=2.8in]{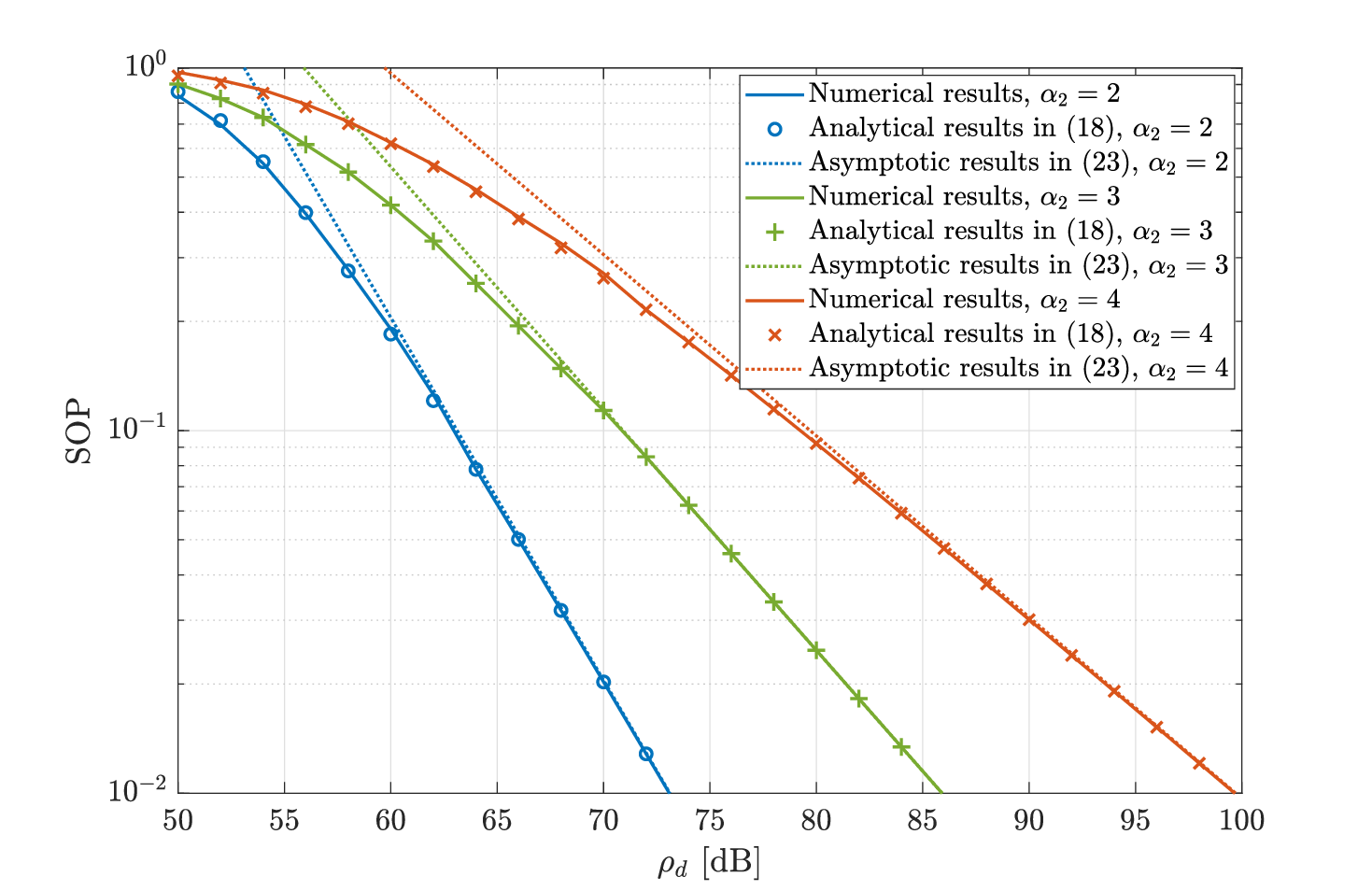}
    \caption{The SOP versus $\rho_d$, with $K=16$, $N=16$, $\alpha_1=2$, $\epsilon=2$, $d_{S\!R}=30~{\rm m}$, $d_{R\!D}=40~{\rm m}$, $r_e=200~{\rm m}$, $\lambda_e=10^{-3}$, $C_{\rm th}=0.05$, and $\rho_e= 60~{\rm dB}$.}
   	\vspace{-0.45cm}
    \label{fig7} \end{figure}

Fig.~\ref{fig7} shows the SOP versus the transmit SNR $\rho_d$ for different values of the path loss exponent $\alpha_2$. It can be seen that the negative slope of secrecy outage curves becomes less steep as $\alpha_2$ increases. Furthermore, the secrecy diversity order presented in \emph{Remark~4} can be verified by calculating the negative slope of the SOP curves on a log-log scale. 

\section{Conclusion}
In this paper, the secrecy performance of an RIS-assisted communication system with randomly located eavesdroppers was studied. The exact distributions of the received SNRs at the legitimate user and the eavesdroppers were presented. Then, closed-form expressions for the SOP and secrecy diversity order were derived. It was demonstrated that the secrecy diversity order primarily depends on the path loss exponent of the RIS-to-ground links. The impact of other key parameters was also analyzed to provide insightful guidelines.

\begin{appendices}
\section{Proof of Theorem 1}
For the transmit beamformer $\mathbf{f}=\frac{\mathbf{g}_D}{\left\|\mathbf{g}_D^H\right\|}$, we compute the optimal reflecting phase shifts at the RIS by maximizing the received signal power as follows
\begin{align}
\mathbf{\Theta}^{\star}\!\!=\!\!\arg\mathop{\max}\limits_{\mathbf{\Theta}}{\!\left|\mathbf{g}_D^H\mathbf{f}\right|^2}\!\mathop=^{\left(\rm d\right)}\!\arg\mathop{\max}\limits_{\mathbf{\Theta}}\!{\left|{\bm{\theta}}^H\!{\rm diag}\!\left\{\!\mathbf{h}_{R\!D}^H\!\right\}\!\mathbf{a}_{N,S\!R}\right|^2},
\label{equal100}
\end{align}
where $({\rm d})$ follows by defining $\bm{\theta}^H\!=\!\!\left[{\rm e}^{j\theta_1},\ldots,{\rm e}^{j\theta_n},\ldots,{\rm e}^{j\theta_N}\right]$. Therefore, the optimal RIS phase shifts are given by
\begin{align}
\theta_n^{\star}=-\angle\left(h_{R\!D}^\ast\left(n\right){\rm a}_{N,S\!R}\left(n\right)\right),
\label{equal101}
\end{align}
and the phase shift matrix can be easily obtained as (\ref{equal8}).

\section{Proof of Lemma 1}
Since $\left|h_{R\!D}\left(1\right)\right|,\left|h_{R\!D}\left(2\right)\right|,\ldots,\left|h_{R\!D}\left(N\right)\right|$ are i.i.d. RVs, the mean and variance of $\left|{\rm A}\right|$ are calculated as $\mathbb{E}\{\left|{\rm A}\right|\}\!\!\!\!=\!\!\!\!\sqrt{K\nu}N\mathbb{E}\{\left|h_{R\!D}\left(n\right)\right|\}$ and ${\rm Var}\{\left|{\rm A}\right|\}=K\nu N{\rm Var}\{\left|h_{R\!D}\left(n\right)\right|\}$, where $\left|h_{R\!D}\left(n\right)\right|\sim Rice\left(\sqrt{\frac{\mu_D\epsilon}{\epsilon+1}}, \sqrt{\frac{1}{2}\frac{\mu_D}{\epsilon+1}}\right)$, and $Rice$ denotes the Rician distribution, whose mean and variance are given as $\mathbb{E}\{\left|h_{R\!D}\left(n\right)\right|\}\!\!=\!\!\sqrt{\frac{\mu_D}{\epsilon+1}}\frac{\sqrt\pi}{2}L_\frac{1}{2}\left(-\epsilon\right)$ and ${\rm Var}\{\left|h_{R\!D}\left(n\right)\right|\}\!=\!\frac{\mu_D}{\epsilon+1}\left[1\!+\!\epsilon\!-\!\frac{\pi}{4}\left(L_\frac{1}{2}\left(-\epsilon\right)\right)^2\right]$, respectively. 

Therefore, according to \cite[Lemma~3]{18}, the RV $\left|{\rm A}\right|$ can be approximated by a Gamma distributed RV with shape parameter $k=\frac{\mathbb{E}\{\left|{\rm A}\right|\}^2}{{\rm Var}\{\left|{\rm A}\right|\}}$ and scale parameter $\theta=\frac{{\rm Var}\{\left|{\rm A}\right|\}}{\mathbb{E}\{\left|{\rm A}\right|\}}$, which yields the desired result in (\ref{equal9}).

\section{Proof of Lemma 2}
From (\ref{equal1}), we see that $h_{Ri}\left(n\right)\!\sim\!\mathcal{CN}\left(\sqrt{\frac{\mu_i\epsilon}{\epsilon+1}}{\overline{h}}_{Ri}\left(n\right),\frac{\mu_i}{\epsilon+1}\right)$, and $\left|h_{Ri}\left(n\right)\right|\!\sim\! Rice\left(\sqrt{\frac{\mu_i\epsilon}{\epsilon+1}},\sqrt{\frac{1}{2}\frac{\mu_i}{\epsilon+1}}\right)$. Then, it can be easily obtained that $\mathbb{E}\{h_{Ri}\!\left(n\right)\}\!=\!\sqrt{\frac{\mu_i\epsilon}{\epsilon+1}}{\rm a}_{N,Ri}\!\left(n\right)$, $\mathbb{E}\{\left|h_{Ri}\!\left(n\right)\right|\}\!=\!\sqrt{\frac{\mu_i}{\epsilon+1}}\frac{\sqrt\pi}{2}L_\frac{1}{2}\!\left(-\epsilon\right)$, and $\mathbb{E}\{\left|h_{Ri}\left(n\right)\right|^2\}\!=\!\mu_i$. It follows that
\begin{align}
\mathbb{E}\{{\rm e}^{-j\angle h_{RD}^\ast\left(n\right)}\}\!=\!\left(\frac{\mathbb{E}\{h_{RD}^\ast\left(n\right)\}}{\mathbb{E}\{\left|h_{RD}^\ast\left(n\right)\right|\}}\right)^\ast\!=\!\frac{\sqrt\epsilon {\rm a}_{N,RD}\left(n\right)}{\frac{\sqrt\pi}{2}L_\frac{1}{2}\left(-\epsilon\right)}.
\label{equal106}
\end{align}

Therefore, the mean and variance of the RV $x_n=h_{RE_m}^\ast\left(n\right){\rm e}^{-j\angle h_{RD}^\ast\left(n\right)}$ can be calculated, respectively, as
\begin{align}
\mathbb{E}\{x_n\}=\sqrt{\frac{\mu_{E_m}\epsilon^2}{\epsilon+1}}\frac{{\rm a}_{N,RE_m}^\ast\left(n\right){\rm a}_{N,RD}\left(n\right)}{\frac{\sqrt\pi}{2}L_\frac{1}{2}\left(-\epsilon\right)},
\label{equal107}
\end{align}
and
\begin{align}
{\rm Var}\{x_n\}=\mu_{E_m}\left[1-\frac{\epsilon^2}{{\frac{\pi}{4}\left(\epsilon+1\right)\left(L_\frac{1}{2}\left(-\epsilon\right)\right)}^2}\right].
\label{equal108}
\end{align}

It can be seen from (\ref{equal107}) that for different $n$, $\mathbb{E}\{x_n\}$ is related to $n$, which means that $x_n$ is not identically distributed. We first define a new RV $x_n-\mathbb{E}\{x_n\}$, and it can be easily verified that $x_1-\mathbb{E}\{x_1\}, x_2-\mathbb{E}\{x_2\}, \ldots, x_N-\mathbb{E}\{x_N\}$ are i.i.d. RVs with zero mean and variance ${\rm Var}\{x_n\}$. By virtue of the central limit theorem (CLT) \cite{250}, $\sum_{n=1}^{N}\left(x_n-\mathbb{E}\{x_n\}\right)$ converges in distribution to a complex Gaussian RV with zero mean and variance $N\!\cdot\!{\rm Var}\{x_n\}$. Then, we can obtain that $Z_{E_m}\sim\mathcal{CN}\left(\sum_{n=1}^{N}\mathbb{E}\{x_n\},N\!\cdot\!{\rm Var}\{x_n\}\right)$, where $\sum_{n=1}^{N}\mathbb{E}\{x_n\}=\sqrt{\frac{\mu_{E_m}\epsilon^2}{{\frac{\pi}{4}\left(\epsilon+1\right)\left(L_{1/2}\left(-\epsilon\right)\right)}^2}}\sum_{0\le x,y\le\sqrt N-1}{\rm e}^{j2\pi\frac{d}{\lambda}\left(x\delta_1+y\delta_2\right)}$ by mapping the index $n$ to the index $\left(x,y\right)$, $\delta_1=\sin{\psi_{RD}^a}\sin{\psi_{RD}^e}-\sin{\psi_{RE_m}^a}\sin{\psi_{RE_m}^e}$, and $\delta_2=\cos{\psi_{RD}^e}-\cos{\psi_{RE_m}^e}$.

\section{Proof of Proposition 2}
Using the asymptotic expansion of the upper incomplete gamma function \cite[Eq.~(6.5.32)]{19}, when $r_e\rightarrow\infty$, the CDF of the overall eavesdropping SNR in (\ref{equal16}) can be given as follows 
\begin{align}
F_{\gamma_E}\left(x\right)\simeq{\rm exp}\left[-t_0x^{-t_4}\Gamma\left(t_1\right)\right].
\label{equal110}
\end{align}
Therefore, the SOP in (\ref{equal18}) is further rewritten as
\begin{align}
{\rm SOP}&=1-\int_{0}^{+\infty}{F_{\gamma_E}\left(\frac{1}{\varphi}\left(1+x\right)-1\right)f_{\gamma_D}\left(x\right)}{\rm{d}}x\nonumber\\
&\simeq1-\frac{1}{2\Gamma\left(k\right)}\left(\sqrt{\rho_d}\theta\right)^{-k}\ I,
\label{equal111}
\end{align}
where $I=\int_{0}^{+\infty}{x^a {\rm exp}\left[-bx^{-c}-\upsilon\sqrt x\right]}{\rm{d}}x$, $a=\frac{k}{2}-1$, $b=t_0\Gamma\left(t_1\right)\varphi^{t_4}$, $c=t_4=\frac{2q}{p}$, and $\upsilon=\frac{1}{\sqrt{\rho_d}\theta}$.

By applying the Mellin convolution theorem \cite{20}, we can get the Mellin transform of $I$ as
\begin{align}
\mathcal{M}\left[I;s\right]=\frac{2p}{2q\upsilon^{2s+2a+2}}\Gamma\left(\frac{ps}{2q}\right)\Gamma\left(2s+2a+2\right).
\label{equal112}
\end{align}
Therefore, we can calculate $I$ using the inverse transform as follows
\begin{align}
I=&\frac{p}{\pi i\upsilon^{2a+2}}\int_{u-i\infty}^{u+i\infty}{\!\Gamma\left(ps\right)\!\Gamma\left(4q\left(s\!+\!\frac{a\!+\!1}{2q}\right)\right)\!\left(\upsilon^{4q}b^p\right)^{-s}}{\rm{d}}s\nonumber\\
\mathop=^{\left(\rm e\right)}&\frac{p^\frac{1}{2}q^{2a+\frac{3}{2}}}{\upsilon^{2a+2}2^{\frac{p+4q}{2}-4a-5}\pi^{\frac{p+4q}{2}-1}}\frac{1}{2\pi i}\int_{u-i\infty}^{u+i\infty}\left(\frac{\upsilon^{4q}b^p}{p^p{256}^qq^{4q}}\right)^{-s}\nonumber\\
&\times\prod_{n=0}^{p-1}\Gamma\left(s+\frac{n}{p}\right)\prod_{n=0}^{4q-1}\Gamma\left(s+\frac{n+2a+2}{4q}\right){\rm{d}}s\nonumber\\
=&\frac{p^\frac{1}{2}q^{2a+\frac{3}{2}}}{\upsilon^{2a+2}2^{\frac{p+4q}{2}-4a-5}\pi^{\frac{p+4q}{2}-1}}G_{0,p+4q}^{p+4q,0}\left(\frac{\upsilon^{4q}b^p}{p^p{256}^qq^{4q}}\middle|\begin{matrix}-\\\Delta\\\end{matrix}\right),
\label{equal113}
\end{align}
where $({\rm e})$ follows from Gauss' multiplication formula \cite[Eq.~(6.1.20)]{19}, and the last equality is derived by applying the definition of Meijer's $G$ function. Subsequently, by substituting (\ref{equal113}) into (\ref{equal111}), the SOP is obtained as shown in (\ref{equal22}).
\end{appendices}

\vspace{12pt}

\end{document}